\documentclass[10pt]{article}
\usepackage{amsmath}
\usepackage{amssymb}
\usepackage{amsthm}
\usepackage{enumerate}
\usepackage{epsf}
\usepackage{psfrag}
\usepackage{mathrsfs}
\usepackage{hyperref}
\usepackage[dvips]{graphicx}
\DeclareGraphicsExtensions{.eps,.art,.ART,.ps}
\newcommand{\cL}{\mathcal{L}}

\newcommand{\bbR}{\mathbb{R}}      

\newcommand{\tr}{\operatorname{tr}}
\newcommand{\cof}{\operatorname{cof}}

\begin{document}
\title{Rigid elastic solids in relativity}
\author{Jos\'e Nat\'ario\\ \\
{\small CAMGSD, Departamento de Matem\'{a}tica, Instituto Superior T\'{e}cnico,}\\
{\small Universidade de Lisboa, Portugal}
}
\date{}
\maketitle
\thispagestyle{empty}
\enlargethispage*{.8cm}
\begin{abstract}
After briefly reviewing the theory of relativistic elasticity,  we expand a general elastic Lagrangian to quadratic order and compute the main parameters for the linear elasticity of relativistic solids: the longitudinal and transverse speeds of sound, the Poisson ratio, and the bulk, shear and Young moduli. Using these, we discuss which Lagrangian is the best choice to model a relativistic rigid elastic solid.
\end{abstract}
\tableofcontents
%
%
\pagebreak
%
%
\section{Introduction}\label{section0}
As is well known, truly rigid solids are incompatible with Einstein's relativity, since no influence can propagate on any given medium faster than the speed of light. The implicit assumption that such solids exist underlies a number of apparent contradictions in the theory of relativity, from Ehrenfest's rotating disk \cite{Ehrenfest09, Gron04} to the car and garage paradox \cite{Rindler61}; these are promptly dismissed once one realizes that all relativistic bodies must be deformable \cite{Brotas68, McCrea71, N14}.

Simple models for deformable solids are provided by the theory of relativistic elasticity, originally formulated by Carter and Quintana \cite{CQ72} (see also \cite{Maugin78, KM92, Tahvildar98, BS03, KS03, Wernig06} and references therein). Such elastic bodies have been studied both in equilibrium configurations \cite{Park00, KS04, FK07, ABS08, ABS09, BCV10, BCMV12, AC14} and in dynamic settings \cite{Magli97, Magli98, CH07, BW07, AOS16, BM17}, and are expected to play an important role in modelling neutron star crusts (see for instance \cite{CH08, AHCS19}).

This paper is concerned with relativistic elastic bodies that, while deformable, are as rigid as possible. In Newtonian mechanics, this would mean making the speeds of sound (or, equivalently, the elastic moduli) tend to infinity, thus obtaining an undeformable body (rigid in the traditional sense); in relativity, however, the speeds of sound cannot be greater than the speed of light, and so a relativistic rigid body should be defined as one where this maximum limit is attained. In fact, there exist a number of elastic materials in the literature with this property: Christodoulou's ``hard phase" fluid \cite{Christodoulou95, CN19}, the stiff ultra-rigid material of Karlovini and Samuelsson \cite{KS04}, or the Brotas solid \cite{Bento85}, to name a few. Our aim here is to decide which (if any) of these materials should be considered the true relativistic version of a rigid solid.

We now describe in detail the results in this paper. In Section~\ref{section0.5} we give a brief overview of the theory of relativistic elasticity, for the convenience of the reader and also to set the notation and conventions. In Section~\ref{section1} we expand the Lagrangian of a  homogeneous and isotropic elastic material to quadratic order about its relaxed configuration and compute the speeds of sound. We define rigid elastic bodies to be those whose {\em longitudinal} speed of sound equals the speed of light (as is the case for all the examples mentioned above; incidentally, their {\em transverse} speeds of sound are all strictly smaller than the speed of light). In Section~\ref{section4} we determine the energy-momentum tensor to linear order, and use it to write the linearized equations of motion and also to compute the Poisson ratio and the elastic moduli. In Section~\ref{section9} we determine the elastic properties of various models of rigid solids in the literature, and also others suggested by our analysis. Finally, we discuss which Lagrangian is the best choice to model a relativistic rigid elastic solid in Section~\ref{section10}, and briefly summarize our findings in Section~\ref{section11}.

We adopt geometrized units, where the speed of light is $c=1$, and follow Einstein's summation convention, with Latin indices representing spatial indices, running from $1$ to $3$, and Greek indices standing for spacetime indices, ranging from $0$ to $3$.
%
%
\section{Relativistic elasticity}\label{section0.5}
In this section we give a brief overview of the theory of relativistic elasticity, for the convenience of the reader and also to set notation and conventions. We follow the discussion in \cite{Bento85} closely.

\begin{figure}[h!]
\begin{center}
\psfrag{M}{$M$}
\psfrag{p}{$\downarrow\pi$}
\psfrag{S}{$S$}
\psfrag{h}{$h$}
\psfrag{k}{$k$}
\epsfxsize=0.6\textwidth
\leavevmode
\epsfbox{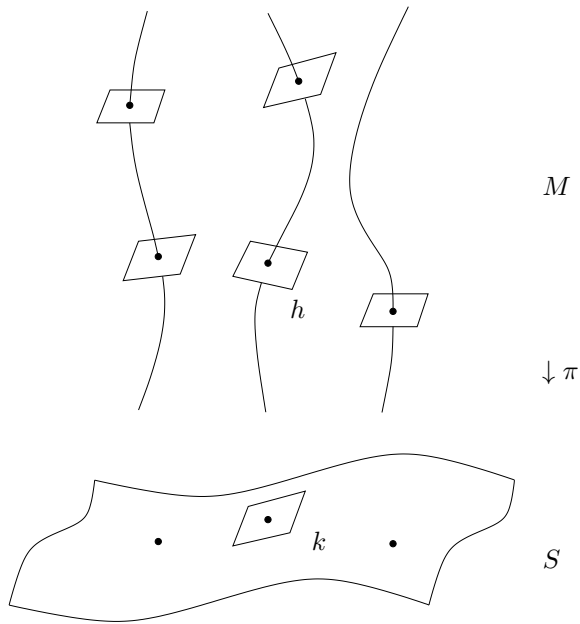}
\end{center}
\caption{A continuous medium in Minkowski's spacetime.} \label{congruence}
\end{figure}

A continuous medium in Minkowski's spacetime\footnote{For simplicity, and also because it suffices for our purposes, we restrict the discussion to elastic materials in Minkowski's spacetime; once the elastic Lagrangian has been obtained, the generalization to arbitrary spacetimes is immediate.} $(M,g)$ can be described by a Riemannian 3-manifold $(S, k)$ (the {\em relaxed configuration}) and projection map $\pi: M \to S$ whose level sets are timelike curves (the worldlines of the medium particles), as shown in Figure~\ref{congruence}.

If we choose local coordinates $(\bar{x}^1, \bar{x}^2, \bar{x}^3)$ on $S$ then we can think of $\pi$ as a set of three scalar fields $\bar{x}^1, \bar{x}^2, \bar{x}^3$ defined on $M$. For a given worldline, we can complete this set of scalar fields into local coordinates $(\bar{t},\bar{x}^1, \bar{x}^2, \bar{x}^3)$ for $M$ such that $\bar{t}$ is the proper time along that worldline and its level sets are orthogonal to it:
\begin{equation}
g = - d\bar{t}^2 + h_{ij} d\bar{x}^i d\bar{x}^j \qquad \text{(on the worldline)}.
\end{equation}
Notice that the orthogonal metric
\begin{equation}
h = h_{ij} d\bar{x}^i d\bar{x}^j
\end{equation}
can be thought of as a time-dependent Riemannian metric on $S$, describing the local deformations of the medium along each worldline, that is, the deviations from the natural metric
\begin{equation}
k = k_{ij} d\bar{x}^i d\bar{x}^j.
\end{equation}
We can compute the (inverse) metric $h$ from
\begin{equation} \label{hij}
h^{ij} = g^{\mu\nu} \frac{\partial \bar{x}^i}{\partial x^\mu} \frac{\partial \bar{x}^j}{\partial x^\nu},
\end{equation}
which does not depend on the choice of $\bar{t}$. In other words, the metric $h$ is a quadratic function of the partial derivatives of the fields $\bar{x}^1, \bar{x}^2, \bar{x}^3$. An {\em elastic} Lagrangian density $\cL$ for these fields is one that depends on the derivatives of the fields only through $h^{ij}$, that is, $\cL = \cL(\bar{x}^i, h^{ij})$. The canonical energy-momentum tensor is
\begin{equation} \label{Tmunu}
T^{\mu\nu} = \frac{\partial\mathcal{L}}{\partial(\partial_\mu \bar{x}^i)} \partial^\nu \bar{x}^i - \mathcal{L} g^{\mu\nu},
\end{equation}
and so in the coordinate system $(\bar{t},\bar{x}^1, \bar{x}^2, \bar{x}^3)$ we have
\begin{equation}
T^{\bar{0}\bar{0}} = \mathcal{L},
\end{equation}
that is, the Lagrangian density is just the rest energy density $\rho = T^{\bar{0}\bar{0}}$ measured by each particle in the medium. The choice of $\rho = \rho(\bar{x}^i, h^{ij})$ is called the {\em elastic law} of the continuous medium.

We define {\em homogeneous and isotropic materials} to be those for which $\rho$ depends only on the eigenvalues $({s_1}^2,{s_2}^2,{s_3}^2)$ of $h_{ij}$ with respect to $k_{ij}$ (that is, the eigenvalues of the matrix $(h_{ij})$ in a frame where $k_{ij}=\delta_{ij}$). Note that $(s_1,s_2,s_3)$ are the dilation factors along the principal directions given by the eigenvectors of $h_{ij}$, that is, the distance between two nearby points along a principal direction in the deformed configuration, as measured by the metric $h$, divided by the distance between the same points in the relaxed configuration, as measured by the metric $k$.

Assume that $k_{ij}=\delta_{ij}$, that is, assume that $(S, k)$ is the Euclidean space. We define the more convenient variables
\begin{align}
& \lambda_0 = \det (h^{ij}) = \frac1{(s_1 s_2 s_3)^2} ; \\
& \lambda_1 = \tr (h^{ij}) = \frac1{{s_1}^2} + \frac1{{s_2}^2} + \frac1{{s_3}^2}; \\
& \lambda_2 = \tr \cof (h^{ij}) = \frac1{(s_1s_2)^2} + \frac1{(s_2 s_3)^2} + \frac1{(s_3 s_1)^2}.
\end{align}
Note that 
\begin{equation}
s_1 s_2 s_3 = \left(\frac1{\lambda_0}\right)^\frac12
\end{equation}
is the ratio between the volume occupied by an element of the material in the deformed state and its volume in the relaxed configuration. Equivalently, 
\begin{equation}
n = (\lambda_0)^\frac12
\end{equation}
is the number density of particles of the medium in the deformed state, if we normalize the number density in the relaxed configuration to be $1$ particle per unit volume. Elastic media whose elastic law depends only on $n$,
\begin{equation}
\rho=\rho(\lambda_0),
\end{equation}
are simply perfect fluids. To check this, we note that from the formula for the inverse of a matrix we have
\begin{equation}
h_{ij} = \frac1{\lambda_0} A^{ji} = \frac1{\lambda_0} A^{ij},
\end{equation}
where $A^{ij}$ is the $(i,j)$-cofactor of $(h^{ij})$. On the other hand, from the Laplace expansion for determinants we have
\begin{equation}
\lambda_0 = \sum_{j=1}^3 h^{ij} A^{ij}  \qquad (\text{no sum over } i),
\end{equation}
and so
\begin{equation}
\frac{\partial \lambda_0}{\partial h^{ij}} = A^{ij} = \lambda_0 h_{ij}.
\end{equation}
Therefore, using \eqref{hij} and \eqref{Tmunu},
\begin{align}
T^{\mu\nu} & = \frac{d \rho}{d \lambda_0} \frac{\partial \lambda_0}{\partial h^{ij}} \frac{\partial h^{ij}}{\partial(\partial_\mu \bar{x}^k)} \partial^\nu \bar{x}^k - \rho g^{\mu\nu} \nonumber \\
& = \frac{d \rho}{d \lambda_0} \lambda_0 h_{ij} (g^{\mu\alpha} \delta^{i}_{\,\,\,\,k} \partial_\alpha \bar{x}^j + g^{\mu\alpha} \partial_\alpha \bar{x}^i \delta^{j}_{\,\,\,\,k})\partial^\nu \bar{x}^k - \rho g^{\mu\nu} \nonumber \\
& = 2\lambda_0\frac{d \rho}{d \lambda_0} h_{ij} \partial^\mu \bar{x}^i \partial^\nu \bar{x}^j - \rho g^{\mu\nu}.
\end{align}
Note that
\begin{equation}
h_{\mu\nu} = h_{ij} \partial_\mu \bar{x}^i \partial_\nu \bar{x}^j
\end{equation}
is simply the metric on the hyperplanes orthogonal to the worldlines, that is,
\begin{equation}
h_{\mu\nu} = g_{\mu\nu} + U_\mu U_\nu,
\end{equation}
where $U$ is the unit tangent vector to the worldlines. Therefore we have
\begin{align}
T^{\mu\nu} & = 2\lambda_0\frac{d \rho}{d \lambda_0}U^\mu U^\nu + \left(2\lambda_0\frac{d \rho}{d \lambda_0} - \rho\right) g^{\mu\nu} \nonumber \\
& = (\rho + p) U^\mu U^\nu + p g^{\mu\nu}
\end{align}
with
\begin{equation}
p = 2\lambda_0\frac{d \rho}{d \lambda_0} - \rho,
\end{equation}
which is indeed the energy-momentum tensor of a perfect fluid. For example, dust corresponds to the elastic law $\rho=\rho_0 \sqrt{\lambda_0}$ (for some positive constant $\rho_0$), yielding $p=0$, and a stiff fluid, with equation of state $p=\rho$, is given by the choice $\rho=\rho_0 \lambda_0$. The rigid elastic fluid considered in \cite{Christodoulou95, CN19}, with equation of state $p=\rho-\rho_0$, corresponds to $\rho=\frac{\rho_0}2(\lambda_0+1)$. 

To obtain elastic materials that are not fluids we must choose elastic laws that also depend on $\lambda_1$ and $\lambda_2$. For instance, an elastic law is said to be {\em quasi-Hookean} if it is of the form
\begin{equation}
\rho = \hat{\rho}(n) + \hat{\mu}(n) \sigma,
\end{equation}
where $\sigma$ is a {\em shear scalar}, that is, a non-negative function of the dilation factors such that $\sigma=0$ if and only if $s_1 = s_2 = s_3$. The functions $\hat{\rho}$ and $\hat{\mu}$ are called the {\em unsheared energy density} and the {\em rigidity modulus} of the elastic material. Examples of these are the {\em John quasi-Hookean material} \cite{Tahvildar98, AC14}, corresponding to the shear scalar
\begin{equation}
\sigma = \frac{{s_1}^2 + {s_2}^2 + {s_3}^2}{\left({s_1}^2 {s_2}^2 {s_3}^2 \right)^\frac13} - 3,
\end{equation}
and the the {\em Karlovini-Samuelsson quasi-Hookean material} \cite{KS03, CH07}, corresponding to the shear scalar
\begin{equation}
\sigma = \frac1{12} \left[ \left( \frac{s_1}{s_2} - \frac{s_2}{s_1} \right)^2 + \left( \frac{s_1}{s_3} - \frac{s_3}{s_1} \right)^2 + \left( \frac{s_2}{s_3} - \frac{s_3}{s_2} \right)^2 \right].
\end{equation}
It is easily seen that the first elastic law is of the form
\begin{equation}
\rho=f(\lambda_0)+g(\lambda_0) \lambda_2,
\end{equation}
whereas the second is of the form
\begin{equation}
\rho=f(\lambda_0)+g(\lambda_0) \lambda_1 \lambda_2.
\end{equation}
Other examples that will be especially important in this work are the {\em stiff ultra-rigid material} of Karlovini and Samuelsson \cite{KS04}, given by
\begin{equation}
\rho=\frac{\rho_0}4 (\lambda_2 + 1),
\end{equation}
and the {\em Brotas rigid solid} \cite{Bento85}, given by
\begin{equation}
\rho=\frac{\rho_0}{8}(\lambda_0 + \lambda_1 + \lambda_2 + 1)
\end{equation}
(where $\rho_0$ is a positive constant).
%
%
\section{Elastic Lagrangian to quadratic order}\label{section1}
We assume that any state where the deformed metric $h$ coincides with the relaxed metric $k$ corresponds to an unstressed equilibrium configuration of our solid. In Minkowski's spacetime, this equilibrium solution is given by
\begin{equation}
\bar{x}^i = x^i.
\end{equation}
We can then consider the linearized solution
\begin{equation}
\bar{x}^i = x^i + \xi^i,
\end{equation}
which can be written out in full as 
\begin{equation}
\begin{cases}
\bar{x}(t,x,y,z)=x+\xi(t,x,y,z) \\
\bar{y}(t,x,y,z)=y+\eta(t,x,y,z) \\
\bar{z}(t,x,y,z)=z+\zeta(t,x,y,z)
\end{cases} .
\end{equation}
From
\begin{equation}
d\bar{x}^i = dx^i + d\xi^i,
\end{equation}
that is,
\begin{equation}
\begin{cases}
d\bar{x}=dx+d\xi \\
d\bar{y}=dy+d\eta \\
d\bar{z}=dz+d\zeta
\end{cases},
\end{equation}
we obtain, to second order on the main diagonal and to first order elsewhere,
\begin{align}
\left(\gamma^{ij}\right) & = \left\langle d\bar{x}^i, d\bar{x}^j \right\rangle  \\
& = \left(
\begin{matrix}
1 + 2 \xi_x + \left\langle d\xi, d\xi \right\rangle & \eta_x + \xi_y & \zeta_x + \xi_z \\
\eta_x + \xi_y & 1 + 2 \eta_y + \left\langle d\eta, d\eta \right\rangle & \zeta_y + \eta_z \\
\zeta_x + \xi_z & \zeta_y + \eta_z & 1 + 2 \zeta_z + \left\langle d\zeta, d\zeta \right\rangle 
\end{matrix}
\right), \nonumber
\end{align}
where $\langle\cdot,\cdot\rangle$ stands for the Minkowski inner product of covectors. Therefore, to second order,
\begin{align}
\lambda_0  = & \,\, \det\left(\gamma^{ij}\right) \nonumber \\
= & \,\, 1 + 2\xi_x + 2\eta_y + 2\zeta_z + 4\xi_x\eta_y + 4\xi_x\zeta_z + 4\eta_y\zeta_z \nonumber \\
& + \left\langle d\xi, d\xi \right\rangle + \left\langle d\eta, d\eta \right\rangle + \left\langle d\zeta, d\zeta \right\rangle \nonumber \\
& - (\eta_x + \xi_y)^2 - (\zeta_x + \xi_z)^2 - ( \zeta_y + \eta_z)^2,
\end{align}
as well as
\begin{align}
\lambda_1  = & \,\, \tr\left(\gamma^{ij}\right) \nonumber \\
= & \,\, 3 + 2\xi_x + 2\eta_y + 2\zeta_z \nonumber \\
& + \left\langle d\xi, d\xi \right\rangle + \left\langle d\eta, d\eta \right\rangle + \left\langle d\zeta, d\zeta \right\rangle
\end{align}
and
\begin{align}
\lambda_2  = & \,\, \tr\cof\left(\gamma^{ij}\right) \nonumber \\
= & \,\, 3 + 4\xi_x + 4\eta_y + 4\zeta_z + 4\xi_x\eta_y + 4\xi_x\zeta_z + 4\eta_y\zeta_z \nonumber \\
& + 2\left\langle d\xi, d\xi \right\rangle + 2\left\langle d\eta, d\eta \right\rangle + 2\left\langle d\zeta, d\zeta \right\rangle \nonumber \\
& - (\eta_x + \xi_y)^2 - (\zeta_x + \xi_z)^2 - ( \zeta_y + \eta_z)^2.
\end{align}
Note that, to this order,
\begin{equation} \label{linear}
\lambda_2 - 3 = (\lambda_0 - 1) + (\lambda_1 - 3)
\end{equation}
and
\begin{align} 
(\lambda_0 - 1)^2 & = (\lambda_1 - 3)^2 = \frac14 (\lambda_2 - 3)^2 = (\lambda_0 - 1)(\lambda_1 - 3) \nonumber \\
& = \frac12 (\lambda_0 - 1)(\lambda_2 - 3) = \frac12 (\lambda_1 - 3)(\lambda_2 - 3). \label{quadratic}
\end{align}
To second order, the elastic Lagrangian is then
\begin{align}
\mathcal{L} = & \,\,a_0 + a_1 (\lambda_0 - 1) + b_1 (\lambda_1 - 3) + c_1 (\lambda_2 - 3) \nonumber \\
& + a_2 (\lambda_0 - 1)^2 + b_2 (\lambda_1 - 3)^2 + c_2 (\lambda_2 - 3)^2 \nonumber \\
& + d_2 (\lambda_0 - 1) (\lambda_1 - 3) + e_2 (\lambda_0 - 1) (\lambda_2 - 3) + f_2 (\lambda_1 - 3) (\lambda_2 - 3),
\end{align}
or, using \eqref{linear} and \eqref{quadratic}
\begin{equation} \label{Lagrangian}
\boxed{\mathcal{L} = a_0 + \tilde{a}_1 (\lambda_0 - 1) + \tilde{b}_1 (\lambda_1 - 3) + \tilde{a}_2 (\lambda_0 - 1)^2} \,\, ,
\end{equation}
where
\begin{align}
& \boxed{\tilde{a}_1 = a_1 + c_1} \,\, , \label{tildea}\\
& \boxed{\tilde{b}_1 = b_1 + c_1} \,\, , \label{tildeb}\\
& \boxed{\tilde{a}_2 = a_2 + b_2 + 4c_2 + d_2 + 2e_2 + 2f_2} \,\, .
\end{align}
%
%
\subsection{Longitudinal waves}\label{section2}
To compute the speed of longitudinal waves we set
\begin{equation} \label{longitudinalansatz}
\begin{cases}
\xi=\xi(t,x) \\
\eta=\zeta=0
\end{cases} .
\end{equation}
In this case we have, to quadratic order,
\begin{equation}
\lambda_0 - 1 = 2\xi_x - \xi_t^2 + \xi_x^2,
\end{equation}
implying
\begin{equation}
(\lambda_0 - 1)^2 = 4\xi_x^2,
\end{equation}
and also
\begin{equation}
\lambda_1 - 3 = 2\xi_x - \xi_t^2 + \xi_x^2.
\end{equation}
Consequently the Lagrangian is
\begin{equation}
\mathcal{L} = a_0 - (\tilde{a}_1 + \tilde{b}_1) \xi_t^2 +  (\tilde{a}_1 + \tilde{b}_1 + 4\tilde{a}_2) \xi_x^2,
\end{equation}
from which one can read off\footnote{In rigor, all one can say is that if solutions of the form \eqref{longitudinalansatz} exist then they must satisfy the one-dimensional wave equation with propagation speed $c_L$; we will confirm in Section~\ref{section4} that such solutions do exist.\label{foot}} the speed $c_L$ of the longitudinal waves:
\begin{equation} \label{longitudinal}
\boxed{c_L^2 = \frac{\tilde{a}_1 + \tilde{b}_1 + 4\tilde{a}_2}{\tilde{a}_1 + \tilde{b}_1}} \,\, .
\end{equation}
%
%
\subsection{Transverse waves}\label{section3}
To compute the speed of transverse waves we set
\begin{equation} \label{transverseansatz}
\begin{cases}
\eta=\eta(t,x) \\
\xi=\zeta=0
\end{cases} .
\end{equation}
In this case we have, to quadratic order,
\begin{equation}
\lambda_0 - 1 = - \eta_t^2,
\end{equation}
implying
\begin{equation}
(\lambda_0 - 1)^2 = 0,
\end{equation}
and also
\begin{equation}
\lambda_1 - 3 = - \eta_t^2 + \eta_x^2.
\end{equation}
Consequently the Lagrangian is
\begin{equation}
\mathcal{L} = a_0 - (\tilde{a}_1 + \tilde{b}_1) \eta_t^2 +  \tilde{b}_1 \eta_x^2,
\end{equation}
from which one can read off\footnote{See footnote \ref{foot}.} the speed $c_T$ of the transverse waves:
\begin{equation} \label{transverse}
\boxed{c_T^2 = \frac{\tilde{b}_1}{\tilde{a}_1 + \tilde{b}_1}} \,\, .
\end{equation}
%
%
\section{Energy-momentum tensor to linear order}\label{section4}
The general formula for the canonical energy-momentum tensor is
\begin{equation}
T^{\mu\nu} = \frac{\partial\mathcal{L}}{\partial (\partial_\mu\bar{x}^i)} \partial^\nu\bar{x}^i - \mathcal{L} g^{\mu\nu}.
\end{equation}
Since
\begin{equation}
\partial_i\bar{x}^j = \delta_{ij} + \partial_i\xi^j,
\end{equation}
we have
\begin{equation}
T^{ij} = \frac{\partial\mathcal{L}}{\partial (\partial_i\xi^j)} + \frac{\partial\mathcal{L}}{\partial (\partial_i\xi^k)} \partial^j\xi^k - \mathcal{L} g^{ij}.
\end{equation}
In particular,
\begin{equation}
T^{xx} = \frac{\partial\mathcal{L}}{\partial \xi_x} + \frac{\partial\mathcal{L}}{\partial (\partial_x\xi^k)} \partial^x\xi^k - \mathcal{L},
\end{equation}
and so, to zeroth order,
\begin{equation}
T^{xx} = 2\tilde{a}_1 + 2\tilde{b}_1 - a_0.
\end{equation}
Since we are assuming that $\xi^i=0$ is an unstressed equilibrium position, we must have
\begin{equation} \label{density}
\boxed{2\tilde{a}_1 + 2\tilde{b}_1 = a_0 = \rho_0} \,\, ,
\end{equation}
where $\rho_0$ is the density of the relaxed configuration. Consequently, we have, to first order,
\begin{equation}
T^{xx} = 2\tilde{a}_1 (\xi_x + \eta_y + \zeta_z) + 2\tilde{b}_1 (\xi_x - \eta_y - \zeta_z) + 8\tilde{a}_2 (\xi_x + \eta_y + \zeta_z).
\end{equation}
Similarly,
\begin{equation}
T^{xy} =  \frac{\partial\mathcal{L}}{\partial \eta_x} + \frac{\partial\mathcal{L}}{\partial (\partial_x\xi^k)} \partial^y\xi^k,
\end{equation}
and so, to first order,
\begin{equation}
T^{xy} = 2\tilde{b}_1 (\xi_y + \eta_x).
\end{equation}
Analogously,
\begin{equation}
T^{xz} = 2\tilde{b}_1 (\xi_z + \zeta_x).
\end{equation}
The remaining components can be computed by analogy:
\begin{align}
& T^{yy} = 2\tilde{a}_1 (\xi_x + \eta_y + \zeta_z) + 2\tilde{b}_1 (- \xi_x + \eta_y - \zeta_z) + 8\tilde{a}_2 (\xi_x + \eta_y + \zeta_z); \\
& T^{zz} = 2\tilde{a}_1 (\xi_x + \eta_y + \zeta_z) + 2\tilde{b}_1 (- \xi_x - \eta_y + \zeta_z) + 8\tilde{a}_2 (\xi_x + \eta_y + \zeta_z); \\
& T^{yz} = 2\tilde{b}_1 (\eta_z + \zeta_y).
\end{align}
These can be summarized as
\begin{equation} \label{Tij}
\boxed{T^{ij} = 2\tilde{b}_1 (\partial_i \xi^j + \partial_j \xi^i) + 2(\tilde{a}_1 - \tilde{b}_1 + 4\tilde{a}_2) (\partial_k \xi^k) \delta_{ij}} \,\, .
\end{equation}
Similarly, we have
\begin{equation}
T^{00} = \frac{\partial\mathcal{L}}{\partial (\partial_0\xi^i)} \partial^0\xi^i - \mathcal{L} g^{00},
\end{equation}
and so, to linear order,
\begin{equation}  \label{T00}
\boxed{T^{00} = \mathcal{L} = a_0 + 2(\tilde{a}_1 + \tilde{b}_1) (\partial_i \xi^i)} \,\, .
\end{equation}
Finally,
\begin{equation}
T^{0i} = \frac{\partial\mathcal{L}}{\partial (\partial_0\xi^j)} (\delta_{ij} + \partial^i\xi^j) - \mathcal{L} g^{0i},
\end{equation}
and so, to linear order,
\begin{equation}  \label{T0i}
\boxed{T^{0i} = \frac{\partial\mathcal{L}}{\partial (\partial_0\xi^j)} = -2(\tilde{a}_1 + \tilde{b}_1) (\partial_0 \xi^i)} \,\, .
\end{equation}
In particular, the linearized equations of motion are
\begin{equation}
\partial_0T^{00} + \partial_iT^{0i} = 0,
\end{equation}
which is easily seen to be automatically satisfied, and
\begin{equation}
\partial_0T^{0i} + \partial_jT^{ij} = 0,
\end{equation}
which can be written in the form
\begin{equation}
\boxed{\partial^2_0 \xi^{i} = c_T^2 \Delta \xi^i + (c_L^2 - c_T^2) \partial_i (\partial_j \xi^j)} \,\, .
\end{equation}
This is the classical Navier-Cauchy equation for linear elasticity of homogeneous and isotropic materials (see for instance \cite{Landau59}), although $\xi^i$ is {\em not} the displacement field (in fact, one can easily show that it is, to linear order, {\em minus} the displacement field). Note that the longitudinal and transverse waves obtained in Section~\ref{section1} are indeed solution of this equation.
%
%
\subsection{Poisson ratio}\label{section5}
The Poisson ratio $\nu$ is the quotient between the infinitesimal dilation along the $xy$-plane (say) and the infinitesimal contraction along the $zz$-axis when the solid is compressed along the $zz$-axis and unconstrained in the $xy$-directions. To compute it we consider the infinitesimal deformation 
\begin{equation} \label{infinitesimal_deformation}
\begin{cases}
\xi = \alpha x \\
\eta = \beta y \\
\zeta = \gamma z
\end{cases} 
\Leftrightarrow
\begin{cases}
x = \bar{x}  - \alpha x \\
y = \bar{y} - \beta y \\
z = \bar{z} - \gamma z
\end{cases}
.
\end{equation}
To first order in $(\alpha,\beta,\gamma)$, we have
\begin{equation}
\begin{cases}
x = (1 - \alpha) \bar{x} \\
y = (1 - \beta) \bar{y} \\
z = (1 - \gamma) \bar{z}
\end{cases}
,
\end{equation}
and so this deformation corresponds to infinitesimal dilations by a factor of $1-\alpha, 1-\beta$ and $1-\gamma$ along the $xx$, $yy$ and $zz$-axes, respectively. Requiring that the pressures along surfaces of constant $xx$ vanish yields
\begin{equation}
(\tilde{a}_1 + 4\tilde{a}_2) (\alpha + \beta + \gamma) + \tilde{b}_1 (\alpha - \beta - \gamma) = 0,
\end{equation}
and requiring that the pressures along surfaces of constant $yy$ vanish yields
\begin{equation}
(\tilde{a}_1 + 4\tilde{a}_2) (\alpha + \beta + \gamma) + \tilde{b}_1 (- \alpha + \beta - \gamma) = 0.
\end{equation}
Therefore, if $\tilde{b}_1 \neq 0$ we must have 
\begin{equation}
\alpha=\beta
\end{equation}
and
\begin{equation}
(2\tilde{a}_1 + 8\tilde{a}_2) \alpha  + (\tilde{a}_1 - \tilde{b}_1 + 4\tilde{a}_2) \gamma = 0. 
\end{equation}
Since the Poisson ratio is the quotient between the infinitesimal increase in the dilation factor along the $xy$-plane, $-\alpha$, and the infinitesimal increase in the contraction factor along the $zz$-axis, $\gamma$, we obtain
\begin{equation} \label{Poisson}
\boxed{\nu = -\frac{\alpha}{\gamma} = \frac{\tilde{a}_1 - \tilde{b}_1 + 4\tilde{a}_2}{2\tilde{a}_1 + 8\tilde{a}_2}} \,\, .
\end{equation}
From equations \eqref{longitudinal}, \eqref{transverse} and \eqref{Poisson} it is easy to check that the classical formula
\begin{equation}
\boxed{\frac{c_T^2}{c_L^2} = \frac{1-2\nu}{2(1-\nu)}}
\end{equation}
still holds for relativistic solids.
%
%
\subsection{Bulk modulus}\label{section6}
The bulk modulus $K$ is the ratio between the infinitesimal pressure applied to a small element of the solid and the infinitesimal relative decrease in its volume. If we set $\alpha=\beta=\gamma$ in \eqref{infinitesimal_deformation} we obtain
\begin{equation}
T^{xx} = T^{yy} = T^{zz} = \left( 6\tilde{a}_1 - 2\tilde{b}_1 + 24\tilde{a}_2 \right) \alpha.
\end{equation}
On the other hand, the infinitesimal relative increase in the volume is
\begin{equation}
(1 - \alpha)^3 - 1 = -3\alpha
\end{equation}
(to linear order in $\alpha$). Since the bulk modulus is the ratio between the infinitesimal isotropic pressure, $T^{xx}$, and {\em minus} the infinitesimal relative {\em increase} in volume, $3\alpha$, it is then
\begin{equation} \label{bulk}
\boxed{K = \frac{T^{xx}}{3\alpha} = 2\tilde{a}_1 - \frac23\tilde{b}_1 + 8\tilde{a}_2} \,\, .
\end{equation}
%
%
\subsection{Shear modulus}\label{section7}
The shear modulus is the ratio between the infinitesimal shear pressure applied to a small element of the solid and the infinitesimal angle by which it is sheared.
If we consider the infinitesimal deformation 
\begin{equation} \label{infinitesimal_deformation_shear}
\begin{cases}
\xi = - \alpha y \\
\eta = 0 \\
\zeta = 0
\end{cases} 
\Leftrightarrow
\begin{cases}
x = \bar{x}  + \alpha y \\
y = \bar{y} \\
z = \bar{z}
\end{cases}
,
\end{equation}
corresponding to a shear deformation in the $xy$-plane by an infinitesimal angle $\alpha$, we obtain
\begin{equation}
T^{xx} = T^{yy} = T^{zz} = T^{xz} = T^{yz} = 0
\end{equation}
and
\begin{equation}
T^{xy} = - 2 \tilde{b}_1 \alpha.
\end{equation}
Since the shear modulus is the ratio between the shear pressure exerted {\em on} the body, $-T^{xy}$ (note that $T^{xy}$ is the shear pressure exerted {\em by} the body), and the infinitesimal shear angle, $\alpha$, it is then
\begin{equation} \label{shear}
\boxed{G = -\frac{T^{xy}}{\alpha} = 2\tilde{b}_1} \,\, .
\end{equation} 
From \eqref{density} we have the classical relation
\begin{equation}
\boxed{\frac{G}{\rho_0} = \frac{\tilde{b}_1}{\tilde{a}_1 +  \tilde{b}_1} = c_T^2} \,\, .
\end{equation}
Similarly, we can also check that
\begin{equation}
\boxed{\frac{K + \frac43 G}{\rho_0} = \frac{\tilde{a}_1 + \tilde{b}_1 + \tilde{a}_2}{\tilde{a}_1 +  \tilde{b}_1} = c_L^2} \,\, ,
\end{equation}
another classical relation.
%
%
\subsection{Young modulus}\label{section8}
The Young modulus is the ratio between the infinitesimal pressure along the $zz$-axis (say) and the infinitesimal contraction along it when the solid is unconstrained in the $xy$-directions. To compute it, we return to the deformation \eqref{infinitesimal_deformation} with $\alpha=\beta$, considered in Section~\ref{section5} when determining the Poisson ratio, and compute the ratio between $T^{zz}$ and the infinitesimal increase in the contraction factor along the $zz$-axis, $\gamma$:
\begin{align}
E & = \frac{T^{zz}}{\gamma} = \left(2\tilde{a}_1 + 8\tilde{a}_2\right) \left(1+\frac{2\alpha}{\gamma}\right) + 2\tilde{b}_1 \left(1-\frac{2\alpha}{\gamma}\right) \nonumber \\
& = \left(2\tilde{a}_1 + 8\tilde{a}_2\right) (1-2\nu) + 2\tilde{b}_1 (1 + 2\nu) \nonumber \\ 
& = K(1-2\nu) + \frac43 G(1 + \nu). 
\end{align}
From \eqref{Poisson}, \eqref{bulk} and \eqref{shear} one can easily show that
\begin{equation}
\boxed{K(1-2\nu) = \frac23 G(1 + \nu)} \,\, ,
\end{equation}
and so we have the classical relations
\begin{equation} \label{Young}
\boxed{E = 3K(1-2\nu) = 2G(1 + \nu)} \,\, .
\end{equation}
%
%
\section{Rigid solids}\label{section9}
We define {\em rigid solids} as those whose longitudinal speed of sound\footnote{We could also consider solids whose transverse speed of sound is as large as possible, $c_T^2=1$, which by \eqref{transverse} is equivalent to choosing $\tilde{a}_1=0$; but then \eqref{longitudinal}, \eqref{bulk} and \eqref{density} show that $c_L^2\leq 1$ implies $K<0$, and so these solids would be unphysical.} is as large as possible, $c_L^2=1$; from \eqref{longitudinal} this is equivalent to choosing $\tilde{a}_2=0$. In this section we will compute the elastic properties of various models of rigid solids in the literature, and also others suggested by our analysis. Each of these examples can be thought of as representing a class of solids, since any Lagrangian which coincides with one of them up to quadratic order will have the same elastic properties.
\subsection{Christodoulou's ``hard phase" material}
In \cite{Christodoulou95}, Christodoulou introduced a two-phase fluid model aiming to describe stellar collapse with the possible formation of a neutron star. This model comprised a dust ``soft phase", representing the uncompressed stellar material, and a rigid fluid ``hard phase", representing stellar material compressed to nuclear densities. The transition between the two phases happened at a certain threshold density, and it was in fact the main focus of research within this model (see \cite{Christodoulou96, Christodoulou96b, CL16}). 

The ``hard phase" material, dubbed a ``rigid elastic fluid" in \cite{CN19}, has also been studied as to the existence of static solutions with a vacuum exterior \cite{FS19} and regarding the propagation of elastic shock waves \cite{CN19}. It is indeed a fluid with equation of state $p=\rho-\rho_0$; as mentioned in Section~\ref{section0.5}, the Lagrangian for this material is an affine function of $\lambda_0$, and so by \eqref{density}
\begin{equation}
\mathcal{L} = a_0 + a_1 (\lambda_0 - 1) = \frac{\rho_0}2 (\lambda_0 + 1).
\end{equation}
Therefore, from \eqref{longitudinal}, \eqref{transverse}, \eqref{Poisson}, \eqref{bulk}, \eqref{shear} and \eqref{Young}, this material has the following elastic characteristics:
\begin{equation}
\begin{cases}
c_L^2 = 1; \\
c_T^2 = 0; \\
\nu = \frac12; \\
K = \rho_0; \\
G = 0; \\
E = 0.
\end{cases} 
\end{equation}
As one would expect from a fluid, there are no transverse sound waves, and both the shear and the Young modulus vanish; therefore this is not a good model for a solid. Note that the Poisson ratio is characteristic of an incompressible material.
\subsection{Sigma model solid}
Another possibility is to take the Lagrangian to be an affine function of $\lambda_1$. This is the Lagrangian for a sigma model consisting of three independent massless scalar fields,
\begin{equation}
\mathcal{L} = \frac12 g^{\mu\nu} \delta_{ij} \partial_\mu \bar{x}^i \partial_\nu \bar{x}^j,
\end{equation}
whose solutions are simply spatial harmonic coordinates. The exact expression of the Lagrangian is, in view of \eqref{density},
\begin{equation}
\mathcal{L} = a_0 + b_1 (\lambda_1 - 3) = \frac{\rho_0}2 (\lambda_1 - 1).
\end{equation}
Therefore, from \eqref{longitudinal}, \eqref{transverse}, \eqref{Poisson}, \eqref{bulk}, \eqref{shear} and \eqref{Young}, this material has the following elastic characteristics:
\begin{equation}
\begin{cases}
c_L^2 = 1; \\
c_T^2 = 1; \\
\nu = \infty; \\
K = - \frac{\rho_0}3; \\
G = \rho_0; \\
E = \infty.
\end{cases} 
\end{equation}
As one would expect, this is a highly rigid material, with both speeds of sound equal to the speed of light; unfortunately, it is also a highly unphysical one, with negative bulk modulus and infinite Poisson ratio and Young modulus.
\subsection{SUREOS material}
A third possibility is to choose a Lagrangian which is an affine function of $\lambda_2$. The corresponding elastic material was studied in \cite{KS04}, where it was dubbed the stiff ultra-rigid equation of state (SUREOS) material. We have, in view of \eqref{density},
\begin{equation}
\mathcal{L} = a_0 + c_1 (\lambda_2 - 3) = \frac{\rho_0}4 (\lambda_2 + 1).
\end{equation}
Therefore, from \eqref{longitudinal}, \eqref{transverse}, \eqref{Poisson}, \eqref{bulk}, \eqref{shear} and \eqref{Young}, this material has the following elastic characteristics:
\begin{equation}
\begin{cases}
c_L^2 = 1; \\
c_T^2 = \frac12; \\
\nu = 0; \\
K = \frac{\rho_0}3; \\
G = \frac{\rho_0}2; \\
E = \rho_0.
\end{cases} 
\end{equation}
From \eqref{Tij}, \eqref{T00} and \eqref{bulk} it is easy to see that, to linear order,
\begin{equation}
T^{00} = \rho_0 + \frac{\rho_0}{3K}\tr (T^{ij}).
\end{equation}
Thus if $p_1$, $p_2$ and $p_3$ are the principal pressures, that is, eigenvalues of $T^{ij}$, then the SUREOS material satisfies $\rho = \rho_0 + p_1 + p_2 + p_3$. Karlovini and Samuelsson argued that in this sense the SUREOS material is similar to the MIT bag (perfect fluid) equation of state $\rho = \rho_0 + 3p$; however, such fluid has very different elastic properties from the SUREOS material, e.g. no transverse sound waves and a much smaller longitudinal speed of sound ($\frac{\sqrt{3}}3$).
\subsection{Brotas rigid solid}

Another rigid solid, studied in \cite{Bento85}, corresponds to the Lagrangian
\begin{equation}
\mathcal{L} = \frac{\rho_0}8 (\lambda_0 + \lambda_1 + \lambda_2 + 1).
\end{equation}
This is an extension of the one-dimensional rigid rods and strings discussed in \cite{HM52, McCrea52, BF03, N14}, based on the idea that the variations in energy density for dilations in orthogonal directions should be independent. From \eqref{longitudinal}, \eqref{transverse}, \eqref{Poisson}, \eqref{bulk}, \eqref{shear} and \eqref{Young}, this material has the following elastic characteristics:
\begin{equation}
\begin{cases}
c_L^2 = 1; \\
c_T^2 = \frac12; \\
\nu = 0; \\
K = \frac{\rho_0}3; \\
G = \frac{\rho_0}2; \\
E = \rho_0.
\end{cases}
\end{equation}
These are exactly the same as those of the SUREOS material, and agree with the transverse speed of sound and the Poisson ratio computed in \cite{Bento85}. In fact, \cite{Bento85} also considered superpositions of the Brotas solid and Christodoulou's rigid elastic fluid,
\begin{equation} \label{alpha1}
\mathcal{L} = \frac{(1-\alpha)\rho_0}8 (\lambda_0 + \lambda_1 + \lambda_2 + 1) + \frac{\alpha \rho_0}2 (\lambda_0 + 1) \qquad (\alpha \in [0,1]),
\end{equation}
corresponding to
\begin{equation} \label{alpha2}
\tilde{a}_1 = \frac{(1+\alpha)\rho_0}4, \qquad \tilde{b}_1 = \frac{(1-\alpha)\rho_0}4.
\end{equation}
The elastic parameters for this solid are
\begin{equation} \label{alpha3}
\begin{cases} 
c_L^2 = 1; \\
c_T^2 = \frac{1-\alpha}2; \\
\nu = \frac{\alpha}{1+\alpha}; \\
K = \frac{(1+2\alpha)\rho_0}3; \\
G = \frac{(1-\alpha)\rho_0}2. \\
E = \frac{(1-\alpha)(1+2\alpha)\rho_0}{1+\alpha}.
\end{cases}
\end{equation}
Again, the transverse speed of sound and the Poisson ratio agree with those computed in \cite{Bento85}.
\subsection{General solids}
Since we defined rigid solids to be those for which
\begin{equation}
c_L^2 = 1 \Rightarrow \tilde{a}_2 = 0,
\end{equation}
such solids are parameterized by pairs $(\tilde{a}_1,\tilde{b}_1)$ satisfying \eqref{density}. If we require a nonnegative bulk modulus then we must have, by \eqref{bulk},
\begin{equation}
K \geq 0 \Rightarrow \tilde{a}_1 \geq \frac13 \tilde{b}_1.
\end{equation}
Using \eqref{density}, this is equivalent to
\begin{equation}
\tilde{a}_1 \geq \frac{\rho_0}8.
\end{equation}
In particular $\tilde{a}_1>0$. Requiring a nonnegative Poisson ratio then implies, by \eqref{Poisson},
\begin{equation}
\nu \geq 0 \Rightarrow \tilde{a}_1 \geq \tilde{b}_1.
\end{equation}
Using \eqref{density}, this is equivalent to
\begin{equation}
\tilde{a}_1 \geq \frac{\rho_0}4.
\end{equation}
Finally, requiring a real transverse speed of sound implies, by \eqref{transverse} and \eqref{density},
\begin{equation}
c_T^2 \geq 0 \Rightarrow \tilde{b}_1 \geq 0.
\end{equation}
Using \eqref{density}, this is equivalent to
\begin{equation}
\tilde{a}_1 \leq \frac{\rho_0}2.
\end{equation}
Therefore we have the range
\begin{equation}
\tilde{a}_1 \in \left[ \frac{\rho_0}4, \frac{\rho_0}2 \right],
\end{equation}
whence
\begin{equation}
\tilde{b}_1 = \frac{\rho_0}2 - \tilde{a}_1 \in \left[ 0, \frac{\rho_0}4 \right] \quad \text{(decreasing)}.
\end{equation}
Correspondingly, we have the following ranges for the various elastic parameters:
\begin{align}
& c_T^2 \in \left[ 0, \frac12 \right] \quad \text{(decreasing)}; \\
& \nu \in \left[ 0, \frac12 \right] \quad \text{(increasing)}; \\
& K \in \left[ \frac{\rho_0}3, \rho_0 \right] \quad \text{(increasing)}; \\
& G \in \left[ 0, \frac{\rho_0}2 \right] \quad \text{(decreasing)}; \\
& E \in \left[ 0, \rho_0 \right] \quad \text{(decreasing)}.
\end{align}
In other words, as $\tilde{a}_1$ increases along its range the elastic properties of the rigid solids interpolate between those of the SUREOS material (or the Brotas solid) and those of Christodoulou's rigid elastic fluid, and are in fact given by equations \eqref{alpha2} and \eqref{alpha3}. From these expressions it is clear that the most rigid solids, that is, those with largest transverse speed of sound, correspond to $\alpha=0$, that is, to
\begin{equation}
\tilde{a}_1 = \tilde{b}_1 = \frac{\rho_0}4.
\end{equation}
From \eqref{tildea} and \eqref{tildeb} it is clear that the corresponding Lagrangian is
\begin{equation} \label{rigidL}
\mathcal{L} = \left( \frac{\rho_0}4 - a_1 \right) + a_1 \lambda_0 + a_1 \lambda_1 + \left( \frac{\rho_0}4 - a_1 \right)\lambda_2,
\end{equation}
with $a_1 \in \bbR$ a free parameter. This includes the SUREOS material ($a_1=0$) and the Brotas solid ($a_1=\frac{\rho_0}8$), among infinitely many other possibilities; the elastic characteristics that we have computed so far are not sufficient to distinguish between these materials.
%
%
\section{Which is the true rigid solid?}\label{section10}
\subsection{Behavior under compression}
To distinguish between the rigid solids given by the family of Lagrangians \eqref{rigidL}, we start with the equilibrium solution given by
\begin{equation}
\bar{x}^i = n_i x^i \qquad \text{(no summation)},
\end{equation}
corresponding a state where the solid is compressed by a different amount along each of the three axes. We can then consider the linearized solution
\begin{equation}
\bar{x}^i = n_i x^i + \xi^i \qquad \text{(no summation)},
\end{equation}
which can be written out in full as 
\begin{equation}
\begin{cases}
\bar{x}(t,x,y,z)=n_1x+\xi(t,x,y,z) \\
\bar{y}(t,x,y,z)=n_2y+\eta(t,x,y,z) \\
\bar{z}(t,x,y,z)=n_3z+\zeta(t,x,y,z)
\end{cases} .
\end{equation}
From
\begin{equation}
d\bar{x}^i = n_idx^i + d\xi^i,
\end{equation}
that is,
\begin{equation}
\begin{cases}
d\bar{x}=n_1dx+d\xi \\
d\bar{y}=n_2dy+d\eta \\
d\bar{z}=n_3dz+d\zeta
\end{cases},
\end{equation}
we obtain, to second order on the main diagonal and to first order elsewhere,
\begin{align}
\left(\gamma^{ij}\right) & = \left\langle d\bar{x}^i, d\bar{x}^j \right\rangle  \\
& = \left(
\begin{matrix}
{n_1}^2 + 2 n_1 \xi_x + \left\langle d\xi, d\xi \right\rangle & n_1\eta_x + n_2\xi_y & n_1\zeta_x + n_3\xi_z \\
n_1\eta_x + n_2\xi_y & {n_2}^2 + 2 n_2 \eta_y + \left\langle d\eta, d\eta \right\rangle & n_2\zeta_y + n_3\eta_z \\
n_1\zeta_x + n_3\xi_z & n_2\zeta_y + n_3\eta_z & {n_3}^2 + 2 n_3 \zeta_z + \left\langle d\zeta, d\zeta \right\rangle 
\end{matrix}
\right). \nonumber
\end{align}
To compute the speed of longitudinal waves we set
\begin{equation}
\begin{cases}
\xi=\xi(t,x) \\
\eta=\zeta=0
\end{cases} .
\end{equation}
In this case we have, to quadratic order,
\begin{align}
\left(\gamma^{ij}\right) = 
\left(
\begin{matrix}
{n_1}^2 + 2 n_1 \xi_x - \xi_t^2 + \xi_x^2 & 0 & 0 \\
0 & {n_2}^2 & 0 \\
0 & 0 & {n_3}^2 
\end{matrix}
\right),
\end{align}
and so
\begin{align}
& \lambda_0 = \left({n_1}^2 + 2 n_1 \xi_x - \xi_t^2 + \xi_x^2\right){n_2}^2{n_3}^2; \\
& \lambda_1 = {n_1}^2 + {n_2}^2 + {n_3}^2 + 2 n_1 \xi_x - \xi_t^2 + \xi_x^2; \\
& \lambda_2 = \left({n_1}^2 + 2 n_1 \xi_x - \xi_t^2 + \xi_x^2\right)\left({n_2}^2 + {n_3}^2\right) + {n_2}^2{n_3}^2.
\end{align}
Substituting into the Lagrangian \eqref{rigidL}, we see that the speed of sound for longitudinal waves will be equal to the speed of light for all our candidate rigid solids. 

To compute the speed of transverse waves we set
\begin{equation}
\begin{cases}
\eta=\eta(t,x) \\
\xi=\zeta=0
\end{cases} .
\end{equation}
In this case we have, to quadratic order,
\begin{align}
\left(\gamma^{ij}\right) =
\left(
\begin{matrix}
{n_1}^2 & n_1\eta_x & 0 \\
n_1\eta_x & {n_2}^2 - \eta_t^2 + \eta_x^2 & 0 \\
0 & 0 & {n_3}^2 
\end{matrix}
\right),
\end{align}
and so
\begin{align}
& \lambda_0 = {n_1}^2{n_2}^2{n_3}^2 - {n_1}^2{n_3}^2 \eta_t^2; \\
& \lambda_1 = {n_1}^2 + {n_2}^2 + {n_3}^2 - \eta_t^2 + \eta_x^2; \\
& \lambda_2 = {n_1}^2{n_2}^2 + {n_1}^2{n_3}^2 + {n_2}^2{n_3}^2 - ({n_1}^2 + {n_3}^2) \eta_t^2 + {n_3}^2\eta_x^2.
\end{align}
Substituting into the Lagrangian \eqref{rigidL}, one can read off the speed $c_T$ of the transverse waves:
\begin{equation} \label{transversecompressed}
c_T^2 = \frac{a_1 + \left( \frac{\rho_0}4 - a_1 \right) {n_3}^2}{a_1 {n_1}^2 {n_3}^2 + a_1 + \left( \frac{\rho_0}4 - a_1 \right) \left({n_1}^2 + {n_3}^2\right)}.
\end{equation}
Note that this speed does not depend on the contraction factor $n_2$ along the $yy$-axis, that is, the direction of oscillation of the transverse waves. It is interesting to note that for the Brotas solid ($a_1=\frac{\rho_0}8$) we have
\begin{equation}
c_T^2 = \frac1{1 + {n_1}^2},
\end{equation}
and so the transverse speed of sound also does not depend on the contraction factor $n_3$ along the $zz$-axis, in agreement with the idea that the energy density variations for dilations in orthogonal directions should be independent for this material.
\subsection{The verdict}
If the material is uniformly compressed, $n_1=n_2=n_3$,  the transverse speed of sound is
\begin{equation}
c_T^2 = \frac{a_1 + \left( \frac{\rho_0}4 - a_1 \right) {n_1}^2}{a_1 {n_1}^4 + a_1 + 2\left( \frac{\rho_0}4 - a_1 \right) {n_1}^2}.
\end{equation}
This quantity will be positive for all $n_1>0$ if and only if
\begin{equation}
a_1 \in \left[ 0, \frac{\rho_0}4 \right],
\end{equation}
and so we exclude all other values of $a_1$ (as these would correspond to unstable compressed configurations). Moreover, as the contraction factor $n_1$ increases, the transverse speed of sound tends to zero except if $a_1=0$, that is, except for the SUREOS material, which satisfies
\begin{equation}
c_T^2 = \frac12
\end{equation}
for all values of the contraction factor. For this reason, the SUREOS material deserves to be considered the true relativistic rigid solid.
%
%
\section{Conclusions}\label{section11}
In this paper we answered the question of which  elastic law that should be regarded as defining the true rigid elastic solid. To do that we expanded the Lagrangian of a  homogeneous and isotropic elastic material to quadratic order about its relaxed configuration and used it to compute its main elastic properties: the longitudinal and transverse speeds of sound, the Poisson ratio and the bulk, shear and elastic moduli. Defining rigid elastic bodies to be those whose longitudinal speed of sound equals the speed of light, and requiring their transverse speed of sound to be maximal, while keeping their elastic properties physical, we were led to a one-parameter family of materials that include the Karlovini-Samuelsson SUREOS material \cite{KS04} and the Brotas solid \cite{Bento85}. To break this degeneracy we considered these materials in a compressed state and found that only one elastic law kept a large transverse speed of sound for all contraction factors, namely the Karlovini-Samuelsson SUREOS material. For this reason, the SUREOS material deserves to be considered the true relativistic rigid solid.
%
%
\section*{Acknowledgements}
This work was partially supported by FCT/Portugal through UID/MAT/04459/2019 and grant (GPSEinstein) PTDC/MAT-ANA/1275/2014.
%
%
%

\end{document}